\documentclass[12pt,english,floatfix,nofootinbib,superscriptaddress,aps,prd,preprint]{revtex4}
\usepackage[utf8]{inputenc}
\usepackage{float}
\usepackage{array}
\usepackage{bbold}
\usepackage{lipsum}
\usepackage{dsfont}
\usepackage{graphicx}
\usepackage{amsmath,amsthm,amsfonts,amssymb}
\usepackage{graphicx}
\usepackage[english]{babel} 
\usepackage{color}
\usepackage{tensor}
\usepackage{esint}
\usepackage[dvips]{epsfig}
\usepackage[dvips]{graphicx}
\usepackage{float}
\usepackage{units}
\usepackage{textcomp}
\usepackage{mathrsfs}
\usepackage{amsmath}
\usepackage[makeroom]{cancel}
\usepackage{amssymb}
\usepackage{amsbsy}
\usepackage{amsfonts}
\usepackage{amssymb,mathrsfs,xcolor}
\usepackage{esint}
\usepackage{braket}
\usepackage{array}
\usepackage{graphicx}

\usepackage{wasysym}
\usepackage{multirow}
\usepackage{wrapfig}
\usepackage{subfig}

\usepackage{stmaryrd}
\usepackage{upgreek}

\makeatletter

\makeatletter\usepackage{babel}

\usepackage{hyperref}
\hypersetup{
    colorlinks,
    citecolor=blue,
    filecolor=green,
    linkcolor=purple,
    urlcolor=red,
}

\usepackage{slashed}

\newcommand{\ie}{\begin{equation}}
\newcommand{\fe}{\end{equation}}
\newcommand{\se}{\begin{eqnarray}}
\newcommand{\ff}{\end{eqnarray}}

\begin{document}

\title{Comment on ``Thermodynamic properties of Schwarzschild black hole in non-commutative gauge theory of gravity''}

\author{A. A. Ara\'{u}jo Filho}
\email{dilto@fisica.ufc.br}

\affiliation{Departamento de Física, Universidade Federal da Paraíba, Caixa Postal 5008, 58051-970, João Pessoa, Paraíba,  Brazil.}

\affiliation{Physics Department, Federal University of Campina Grande, Caixa Postal 10071, 58429-900, Campina Grande-PB, Brazil}


\author{Iarley P. Lobo}
\email{lobofisica@gmail.com}

\affiliation{Department of Chemistry and Physics, Federal University of Para\'iba, Rodovia BR 079 - km 12, 58397-000 Areia-PB,  Brazil}

\affiliation{Physics Department, Federal University of Campina Grande, Caixa Postal 10071, 58429-900, Campina Grande-PB, Brazil}


\date{\today}

\begin{abstract}

A recent study [Annals Phys. 455 (2023) 169394, e-Print: 2204.01901 [gr-qc]] examined the thermodynamic behavior of {an axially symmetric} 
black hole within a non--commutative framework { that mimics the effect of an angular momentum}. However, the analysis presents notable computational inconsistencies. In that analysis, the event horizon was miscalculated, and this error propagated through and compromised all subsequent results. In addition, an incorrect definition of surface gravity was used—the {spherically symmetric} case was invoked for {an axially symmetric} spacetime—rendering the thermodynamic results invalid. In other words, {all the results presented in the paper require a thorough reexamination.}

\end{abstract}

\maketitle


{\section{The black hole solution: the general features}}

The work titled “Thermodynamic properties of Schwarzschild black hole in non--commutative gauge theory of gravity” \cite{Touati:2022zbm} investigates thermodynamical features of {an axially symmetric black hole} derived from a non--commutative gauge theory framework. Such a methodology of implementing non--commutativity has been used in a variety of contexts \cite{Juric:2025kjl,araujo2024effects,araujo2025properties,heidari2025non,heidari2025noaan}.  Nevertheless, several computational inconsistencies are present in the analysis. The purpose of this comment is to point out these problems.

Essentially, the analysis in that work relies on a specific choice for the non–commutative matrix, namely $\Theta = \Theta^{42} = -\Theta^{24}$, namely, 
\begin{equation}
	\Theta^{\mu\nu}=\left(\begin{matrix}
		0	& 0 & 0 & 0 \\
		0	& 0 & 0 & \Theta  \\
		0	& 0 & 0 & 0 \\
		0	& -\Theta & 0 & 0
	\end{matrix}
	\right), \qquad \mu,\nu=0,1,2,3,
\end{equation}
for the sake to introduce the non--commutative structure, generating the corresponding radial component
\begin{equation}
\begin{split}
	g_{rr}= & \left(1-\frac{2 M}{r}\right)^{-1} \\
    & +\left\{\frac{M\left(12M^2+Mr\left(-14+\sqrt{1-\frac{2M}{r}}\right)-r^2\left(5+\sqrt{1-\frac{2M}{r}}\right)\right)}{8r^2(-2M+r)^3}\right\}\Theta^{2} \sin^{2}\theta,\label{radial}\\
\end{split}
\end{equation}
which, by itself, was sufficient to carry out the subsequent calculations, specifically those concerning the thermal properties.

According to the authors, evaluating the solution obtained from $1/g_{rr}$ allowed one to determine the position of the event horizon, which they expressed in \cite{Touati:2022zbm} as:
\begin{equation}
\label{wrongevent}
	r_{h}^{NC}=r_{h}\left[ 1+\frac{3}{8}\left( \frac{\Theta }{r_{h}}\right)^{2} \sin^{2}\theta \right],
\end{equation}
in which $r_{h} = 2M$. {\bf{The result stated above is incorrect}}. {As it is straightforward to verify, the divergence occurs only at a single point, $r = 2M$, which coincides with the result obtained for the Schwarzschild case. 
Moreover, in the same reference, the authors derived a perturbed expression for the event horizon, incorporating corrections arising from $\Theta$. In what follows, we shall examine this claim.

To verify such a claim, let us instead} carry out a series expansion of $1/g_{rr}$ assuming $\Theta$ to be small. This procedure yields:
\ie
\label{gasdsad}
\frac{1}{g_{rr}} \approx \left(1-\frac{2 M}{r}\right) + \frac{\Theta ^2 M \sin^{2}\theta \left(12 M^2-r^2 \sqrt{\frac{r-2 M}{r}}-14 M r+M r \sqrt{\frac{r-2 M}{r}}-5 r^2\right)}{8 r^4 (2 M-r)}.
\fe
This leads to ten lengthy solutions. To obtain a compact expression for the event horizon—analogous to Eq. (\ref{wrongevent}) in \cite{Touati:2022zbm}—these ones {\bf{are expanded up to second order in $\Theta$}}, resulting in $r_i$ ($i = 1, \ldots, 10$)\footnote{{Note that if the solution of Eq.~(\ref{gasdsad}) were considered without expanding the results, it would lose physical consistency, as it would include higher--order terms in $\Theta$ (e.g., $\Theta^{4}$), which exceed the accuracy of the black hole solution, valid only up to $\Theta^{2}$.}}:
\ie
\begin{split}
r_{1} = &  \frac{i \Theta  \sin (\theta )}{24 \sqrt{3}}+\frac{3463 (-1)^{5/6} \Theta ^{5/3} \sin ^{\frac{5}{3}}(\theta )}{746496 \sqrt[6]{3} M^{2/3}}-\frac{71 \Theta ^{4/3} \left((-1)^{2/3} \sin ^{\frac{4}{3}}(\theta )\right)}{1728 \left(\sqrt[3]{3} \sqrt[3]{M}\right)} \\
& -\frac{1}{2} \Theta ^{2/3} \left(\sqrt[3]{-3} \sqrt[3]{M \sin ^2(\theta )}\right)-\frac{5 \Theta ^2 \sin ^2(\theta )}{48 M},
\end{split}
\fe
\ie
\begin{split}
r_{2} = & \, \,  \Theta ^{2/3} \sqrt[3]{M \sin ^2(\theta )} \alpha_{i} -\frac{i \Theta  \sin (\theta )}{24 \sqrt{3}}\\
& -\frac{3463 \Theta ^{5/3} \left(\sqrt[6]{-\frac{1}{3}} \sin ^{\frac{5}{3}}(\theta )\right)}{746496 M^{2/3}}  +\frac{71 \sqrt[3]{-\frac{1}{3}} \Theta ^{4/3} \sin ^{\frac{4}{3}}(\theta )}{1728 \sqrt[3]{M}}-\frac{5 \Theta ^2 \sin ^2(\theta )}{48 M},
\end{split}
\fe
\ie
\begin{split}
r_{3} = & \, \, \Theta ^{2/3} \sqrt[3]{M \sin ^2(\theta )} \alpha_{i}+\frac{i \Theta  \sin (\theta )}{24 \sqrt{3}}\\
& +\frac{3463 \sqrt[6]{-\frac{1}{3}} \Theta ^{5/3} \sin ^{\frac{5}{3}}(\theta )}{746496 M^{2/3}}  +\frac{71 \sqrt[3]{-\frac{1}{3}} \Theta ^{4/3} \sin ^{\frac{4}{3}}(\theta )}{1728 \sqrt[3]{M}}-\frac{5 \Theta ^2 \sin ^2(\theta )}{48 M},
\end{split}
\fe
\ie
\begin{split}
r_{4} = \, & \Theta ^{2/3} \sqrt[3]{M \sin ^2(\theta )} \alpha_{i}+\frac{i \Theta  \sin (\theta )}{24 \sqrt{3}} \\
& +\frac{3463 \sqrt[6]{-\frac{1}{3}} \Theta ^{5/3} \sin ^{\frac{5}{3}}(\theta )}{746496 M^{2/3}}  +\frac{71 \sqrt[3]{-\frac{1}{3}} \Theta ^{4/3} \sin ^{\frac{4}{3}}(\theta )}{1728 \sqrt[3]{M}}-\frac{5 \Theta ^2 \sin ^2(\theta )}{48 M},
\end{split}
\fe
\ie
\begin{split}
r_{5} = & -\frac{i \Theta  \sin (\theta )}{24 \sqrt{3}}+\frac{3463 i \Theta ^{5/3} \sin ^{\frac{5}{3}}(\theta )}{746496 \sqrt[6]{3} M^{2/3}}-\frac{71 \Theta ^{4/3} \sin ^{\frac{4}{3}}(\theta )}{1728 \left(\sqrt[3]{3} \sqrt[3]{M}\right)}\\
& +\frac{1}{2} \sqrt[3]{3} \Theta ^{2/3} \sqrt[3]{M \sin ^2(\theta )}-\frac{5 \Theta ^2 \sin ^2(\theta )}{48 M},
\end{split}
\fe
\ie
\begin{split}
r_{6} = & -\frac{\Theta  \sqrt{-\sin ^2(\theta )}}{24 \sqrt{3}}+\frac{3463 \Theta ^{5/3} \sqrt{-\sin ^2(\theta )} \sqrt[3]{\sin ^2(\theta )}}{746496 \sqrt[6]{3} M^{2/3}}-\frac{71 \Theta ^{4/3} \sin ^2(\theta )^{2/3}}{1728 \left(\sqrt[3]{3} \sqrt[3]{M}\right)} \\
& +\frac{1}{2} \sqrt[3]{3} \Theta ^{2/3} \sqrt[3]{M \sin ^2(\theta )}-\frac{5 \Theta ^2 \sin ^2(\theta )}{48 M},
\end{split}
\fe
\ie
\begin{split}
r_{7} = & \, \, 2 M-\frac{3}{4} i \Theta  \sin (\theta )-\frac{\Theta ^{3/2} \sqrt{\frac{i \sin ^3(\theta )}{M}}}{32 \sqrt{6}}+\frac{5 \Theta ^2 \sin ^2(\theta )}{32 M},
\end{split}
\fe
\ie
\begin{split}
r_{8} = & \, \, 2 M-\frac{3}{4} i \Theta  \sin (\theta )-\frac{\Theta ^{3/2} \sqrt{\frac{i \sin ^3(\theta )}{M}}}{32 \sqrt{6}}+\frac{5 \Theta ^2 \sin ^2(\theta )}{32 M},
\end{split}
\fe
\ie
r_{9} = \, \, 2 M+\frac{3}{4} i \Theta  \sin (\theta )+\frac{\Theta ^{3/2} \sin ^{\frac{3}{2}}(\theta )}{32 \sqrt{6} \sqrt{i M}}+\frac{5 \Theta ^2 \sin ^2(\theta )}{32 M},
\fe
and
\ie
r_{10} = \, \, 2 M+\frac{3}{4} i \Theta  \sin (\theta )+\frac{\Theta ^{3/2} \sin ^{\frac{3}{2}}(\theta )}{32 \sqrt{6} \sqrt{i M}}+\frac{5 \Theta ^2 \sin ^2(\theta )}{32 M},
\fe
where $\alpha_{i}$ is a complex number.

As an illustration, when setting $M = 1$, $\theta = \pi/2$, and $\Theta = 0.01$, none of the obtained roots are both real and positive, as shown below:
$r_{1} = -0.0167171 - 0.028799 i$,
$r_{2} = -0.0167171 + 0.028799 i$,
$r_{3} = -0.016714 + 0.0292819 i$,
$r_{4} = -0.016714 + 0.0292819 i$,
$r_{5} =  0.0333999 - 0.00023877 i$,
$r_{6} =  0.0333999 - 0.00023877 i$,
$r_{7} =  2.00001 - 0.00750902 i$,
$r_{8} =  2.00001 - 0.00750902 i$,
$r_{9} =  2.00002 + 0.00749098 i$,
$r_{10} = 2.00002 + 0.00749098 i$.
{\bf{The fact that all solutions are either negative or complex is sufficient to invalidate the event horizon radius}} given in Eq. \eqref{wrongevent}. 

Moreover, imposing the condition $\theta = \pi/2$ for $r_{10}$, it leads to the result found in \cite{araujo2025comment}:
\ie
\label{truehorizon}
r_{10} = 2 M+\frac{3}{4} i \Theta  +\frac{\Theta^{3/2}}{32 \sqrt{6} \sqrt{i M}}+\frac{5 \Theta ^2 }{32 M}.
\fe

A straightforward comparison between the outcomes (the event horizons) obtained in Ref. \cite{Touati:2022zbm} and those presented in this work, namely,
\ie
\nonumber
r_{h} = \underbrace{2M\left[ 1+\frac{3}{8}\left( \frac{\Theta }{2M}\right)^{2} \sin^{2}\theta \right]}_{\text{Ref.}\, \cite{Touati:2022zbm} }  \quad\quad \text{and} \quad\quad r_{h} = \underbrace{2 M+\frac{3}{4} i \Theta  \sin (\theta )+\frac{\Theta ^{3/2} \sin ^{\frac{3}{2}}(\theta )}{32 \sqrt{6} \sqrt{i M}}+\frac{5 \Theta ^2 \sin ^2(\theta )}{32 M}}_{\text{This paper}}
\fe
shows that certain additional terms were omitted in their analysis, resulting in contributions that are either negative or imaginary. {Therefore, the claims in Ref.~\cite{Touati:2022zbm} {\bf{fail on two fronts}}: (i) the event horizon admits no $\Theta$--dependent correction; and (ii) even after a perturbative expansion of $1/g_{rr}$, the resulting expressions remain incorrect as demonstrated in this Comment.}

Consequently, because the thermodynamic analysis relies directly on the correct determination of the event horizon, every calculation in the cited work requires revision based on the proper horizon expression. However, even after applying the aforementioned corrections, the physical relevance of the thermal treatment remains questionable, as all roots obtained from the expansion of $1/g_{rr}$ ($r_{1}$ to $r_{10}$) exhibit either negative values or non–zero imaginary components (at least for the value of parameters considered here).

{In addition, the authors evaluated the Hawking temperature, along with other related thermal quantities, using the surface gravity but adopted an inappropriate expression. They applied the formula valid for static and spherically symmetric black hole, whereas the metric in question described a static and axially symmetric spacetime, whose surface gravity should be calculated from its time-like Killing vector \cite{Ridgway:1995ke}.} As a result, together with the incorrectly determined event horizon, their analysis yields flawed expressions for the thermodynamic quantities.


\section*{Acknowledgments}
\hspace{0.5cm}

A. A. Araújo Filho acknowledges support from the Conselho Nacional de Desenvolvimento Científico e Tecnológico (CNPq) and the Fundação de Apoio à Pesquisa do Estado da Paraíba (FAPESQ) under grant [150891/2023-7]. I. P. L. was partially supported by the National Council for Scientific and Technological Development - CNPq grant 312547/2023-4. I. P. L. would like to acknowledge networking support by the COST Action BridgeQG (CA23130) and by the COST Action RQI (CA23115), supported by COST (European Cooperation in Science and Technology). The authors also express gratitude to N. Heidari for the useful discussions. {Moreover, all calculations in this paper were performed using \textit{Mathematica} to carry out the required symbolic computations accurately.}

\section{Data Availability Statement}

Data Availability Statement: No Data associated in the manuscript


\bibliographystyle{ieeetr}
\bibliography{main}

\end{document}